\documentclass[%
reprint,
superscriptaddress,
 amsmath,amssymb,
pra,
]{revtex4-1}
\usepackage[utf8]{inputenc}
\usepackage{gensymb}
\usepackage{graphicx}% Include figure files
\usepackage{dcolumn}% Align table columns on decimal point
\usepackage{bm}% bold math
\usepackage{mathtools}
\usepackage{epstopdf} % .eps figures
\usepackage{color}

\newcommand{\eq}[1]{Eq.~(\ref{#1})}

\newcommand{\sect}[1]{Section \ref{sect:#1}}

\newcommand{\olc}[1]{Ref.~\onlinecite{#1}}

\newcommand{\ket}[1]{| #1 \rangle}
\newcommand{\bra}[1]{\langle #1 |}

\newcommand{\product}[2]{\langle #1 | #2 \rangle}

\begin{document}

%\preprint{APS/123-QED}

\title{Vortex line in the unitary Fermi gas}

\author{Lucas Madeira}
\email{lucas.madeira@asu.edu}
\affiliation{Instituto de F\'{i}sica Gleb Wataghin, Universidade Estadual de Campinas, UNICAMP, Campinas 13083-859, Brazil}
\affiliation{Department of Physics and Astronomy, Arizona State University, Tempe, Arizona 85287, USA}

\author{Silvio A. Vitiello} 

\affiliation{Instituto de F\'{i}sica Gleb Wataghin, Universidade Estadual de Campinas, UNICAMP, Campinas 13083-859, Brazil}

\author{Stefano Gandolfi}
\affiliation{Theoretical Division, Los Alamos National Laboratory, Los Alamos, New Mexico 87545, USA}

\author{Kevin E. Schmidt}%
\affiliation{Department of Physics and Astronomy, Arizona State University, Tempe, Arizona 85287, USA}

\date{\today}% It is always \today, today,
             %  but any date may be explicitly specified

\begin{abstract}
We report diffusion Monte Carlo results for the ground state of
unpolarized spin-1/2 fermions in a cylindrical container and
properties of the system with a vortex-line excitation. The density
profile of the system with a vortex line presents a non-zero density
at the core. We calculate the ground-state energy per particle, the
superfluid pairing gap, and the excitation energy per particle. These
simulations can be extended to calculate the properties of vortex
excitations in other strongly interacting systems, such as
superfluid neutron matter using realistic nuclear Hamiltonians.
\end{abstract}

 %Fermi gas, Fermions systems (quantum statistical mechanics)

%\pacs{Valid PACS appear here}% PACS, the Physics and Astronomy
                             % Classification Scheme.
%\keywords{Suggested keywords}%Use showkeys class option if keyword
                              %display desired
\maketitle

%\tableofcontents

\section{Introduction}
\label{sec:intro} 

Ultracold Fermi gases are dilute systems with interparticle
interactions that can be controlled through Feshbach resonances, which allow
the access of strongly interacting regimes. Until recently, superfluids were
classified as either Bardeen-Cooper-Schrieffer (BCS) states or Bose-Einstein condensates (BECs). In fact they are limit cases of
a continuum of the interaction strength. The possibility of tuning the
parameters to observe changes from one regime to the
other is conceptually interesting, but real enthusiasm came from the
experimental realization of the BCS-BEC crossover \cite{reg04}.% In the
%middle of the crossover lies a strongly interacting system known as the
%unitary Fermi gas, with remarkable properties.

The three-dimensional unitary Fermi gas is a strongly interacting system
with short-range interactions of remarkable properties. When the scattering length $a$
diverges, $1/a k_F \to 0$ ($k_F$ is the Fermi momentum of the system),
the low-energy $s$-wave scattering phase shift is $\delta_0=\pi/2$. The
ground-state energy per particle $E_0$ is proportional to that of the
noninteracting Fermi gas $E_{FG}$ in a box,
\begin{equation}
\label{eq:fg}
E_0=\xi E_{FG} = \xi \frac{3}{10} \frac{\hbar^2 k_F^2}{M},
\end{equation} 
where the constant $\xi$ is known as the Bertsch parameter and $M$
is the mass of the fermion. In the limit $a k_F\rightarrow - \infty$,
quantum Monte Carlo (QMC) results give the exact value of $\xi=0.372(5)$
\cite{car11}, in agreement with experiments \cite{ku12,zur13}.

One signature of superfluidity is the formation of quantized
vortices. Since their first observations in superfluid $^4$He a large
body of experimental and theoretical work has been carried out concerning
bosonic systems \cite{don91,vit96,Ortiz:1995,gio96}. On the other hand, the discovery of
vortex lattices in a strongly interacting rotating Fermi gas of $^6$Li
\cite{zwi05} was a milestone in the study of superfluidity in cold
Fermi gases.

A vortex line consists of an extended irrotational flow field, with a core
region where the vorticity is concentrated. The quantization of the flow
manifests itself in the quantized units $h/2M$ of circulation.
There is no evidence of quantized vortices with more than one unit
of circulation. Many questions remain to be answered concerning the
structure of the vortex core for fermions.

In this paper we focus on ultracold Fermi gases, but our results are useful
also to understand the properties of related systems. Ultracold atomic gases and
low-density neutron matter are unique in the sense that both exhibit
pairing gaps of the order of the Fermi energy \cite{bro13}. The
neutron scattering length is about $-$18.5 fm, which is significantly
larger than the interparticle spacing and the interaction range 2.7 
fm~\cite{Gezerlis:2008}, therefore
low-density neutron matter is also near unitarity. In this regime both
dilute cold fermion
atoms and neutron matter have similar properties~\cite{Carlson:2012}.
The possibility of tuning
particle-particle interactions experimentally in cold
atomic gases provides an emulation of low-density neutron matter,
which is beyond direct experimental reach. We present results
of the vortex structure in cold atomic gases, which can be extended to
direct simulations of vortices in superfluid neutron matter using
realistic nuclear Hamiltonians.

Here we report results for a single vortex line in a unitary
Fermi gas in a cylindrical geometry. We found that the density profile
is flat at the center of the cylinder. % and vanishes smoothly at the wall.
We separated the wall contribution from the ground-state of the system, and determined an upper bound of the
bulk energy as $E_0=(0.50 \pm 0.01) E_{FG}$ per particle. 
$E_{FG}$ is the free Fermi gas energy in the same geometry, defined as
\begin{equation}
E_{FG}=\frac{3}{10} \frac{\hbar^2}{M} \left( 3\pi^2 \frac{A}{V} \right)^{2/3} \,,
\end{equation}
where $A$ is the number of particles, and $V$ is the volume.
 We
also estimated an upper-bound value of the superfluid pairing gap for
this geometry, $\Delta=(1.12\pm 0.02)E_{FG}$. For the system with
a vortex line we obtained the density profile with a nonzero density at the core, and an excitation energy of
$E_{ex}=(8.6\pm0.3)10^{-3}E_{FG}$ per particle.

We have organized this work as follows. In \sect{met} we present the methods employed to treat our system. We begin with
the solution of Schroedinger's equation for a spinless free particle in a
cylindrical container. In \sect{bcs} we show that the
component of the BCS wave function with a fixed number of particles can
be written as an antisymmetrized product of pairing functions, determined for the cylindrical geometry. We introduce the ground-state
wave function and the wave function for the system with a vortex line
in Sections \ref{sect:ground} and \ref{sect:vortex}, respectively. The
QMC methods we employed are briefly described in \sect{qmc}. \sect{res}
contains our results. First, we present the
spatial distribution of the particles in the cylinder for both the ground-state and the system with a vortex line. Energy related quantities,
such as the ground-state energy, the superfluid pairing gap, and the
vortex excitation energy are given in Sec. \ref{sec:energy}.  Finally,
we discuss our results in \sect{con}.

\section{Methods}
\label{sect:met} 

Previous calculations on Bose systems like $^4$He have often used a
periodic array of counter rotating vortices in order to have periodic
boundary conditions to minimize surface effects. For example, in $^4$He
the calculations in \olc{sad97} used 300 particles and 4 counter-rotating vortices in the simulation cell. In order to use the same
number of fermion pairs we would require a system of 600 fermions. While
a few simulations of fermions have been performed with this number of
particles, the required variance for a detailed optimization is beyond
the goals of this paper. We used a circular cylindrical simulation
cell of radius $\mathcal{R}$ with hard wall boundary conditions, at this
radius. The system is periodic in the axial direction.

We begin our calculations by investigating properties of
the ground state of the system. The model
we considered consists of $A$ spin-1/2 fermions in a cylinder of radius
${\cal R}$ and height ${\cal L}$.
Because the system is dilute, the interaction is $s$-wave. Fermions
of the same spin do not interact and we use a short-range potential that
is attractive, which can reproduce the regime of $ak_F \to -\infty$.
The $s$-wave potential $V(r)$ acting between particles with opposite
spin has the form
\begin{equation}
V(r) = -v_0 \frac{2\hbar^2}{M} \frac{\mu^2}{\cosh^2{(\mu r)}},
\end{equation}
where $v_0$ can be adjusted to tune the value of $ak_F$ and $\mu$
controls the effective range $r_{eff}$ of the potential. We set $v_0=1$, which, for this potential, corresponds to $a=\pm \infty$ and $r_{eff}=2/\mu$
\cite{car03,cha04}. In this paper all the calculations are performed
with $\mu r_0=$ 24, and $4\pi r_0^3 n_0 =3$, where $n_0$ is the number
density, but the results could be straightforwardly extrapolated as done in~\olc{Forbes:2012}.

\subsection{Schroedinger's equation in cylindrical coordinates}
\label{sect:cylindrical}

We considered the free-particle solution of Schroedinger's equation
in a cylinder of radius ${\cal R}$ and height ${\cal L}$, finite at $\rho=0$, and subject to
the boundary conditions
\begin{eqnarray}
&\Psi_{nmp}(\rho={\cal R},\varphi,z) = 0,& \nonumber  \\
&\Psi_{nmp}(\rho,\varphi,z) = \Psi_{nmp}(\rho,\varphi+2\pi,z),& \nonumber \\
&\Psi_{nmp}(\rho,\varphi,z) = \Psi_{nmp}(\rho,\varphi,z+\mathcal{L}),&
\end{eqnarray}
where $(\rho,\varphi,z)$ are the usual cylindrical 
coordinates. %; $\rho$ is the distance from the cylinder axis, $\varphi$ is the angle between the Cartesian unit vector $\hat{\textbf{x}}$ and $\hat{\boldsymbol{\rho}}$, and $z$ 
%is the component along the height of the cylinder.
The solution is given by
\begin{equation}
\label{eq:sch_cyl}
\Psi_{nmp}(\rho,\varphi,z) = \mathcal{N}_{mp} J_m\left(k_{mp} \rho\right) \exp{\left[i ( k_z z+m \varphi)\right]},
\end{equation}
where $\mathcal{N}_{mp}$ is a normalization constant, $J_m(k_{mp}
\rho)$ are Bessel functions of the first kind, $k_{mp}=j_{mp}/\mathcal{R}$,
$j_{mp}$ is the $p$-th zero of $J_m$, and $k_z=2\pi n/{\cal L}$. The
quantum numbers $n$ and $m$ can take the values $0,\pm 1, \pm 2, \dots$
and $p=1,2,\dots$ . The eigenvalues of these functions are
\begin{equation}
\label{eq:eigenenergies}
E_{nmp} = \frac{\hbar^2}{2M} \left[ \left(\frac{j_{mp}}{{\cal R}}\right)^2 + \left( \frac{2 \pi n}{\mathcal{L}} \right)^2 \right].
\end{equation}
The set of states $\{ \Psi_{nmp} \}$ is complete, therefore it is
used to expand our many-body trial wave function.

\subsection{BCS wave function projected to a fixed number of particles}
\label{sect:bcs}

The BCS wave function used to describe the Cooper pairs in
the ground-state is written as
\begin{eqnarray}
\label{eq:BCS_theta}
&| BCS \rangle_\theta = \prod\limits_{\textbf{k}} (u_{\textbf{k}}+e^{i\theta} v_{\textbf{k}} \hat{a}^\dagger_{\textbf{k} \uparrow} \hat{a}^\dagger_{-\textbf{k} \downarrow})|0\rangle ,& \nonumber\\
& u_{\textbf{k}}^2+v_{\textbf{k}}^2=1,&
\end{eqnarray}
where $u_{\textbf{k}}$ and $v_{\textbf{k}}$ are real numbers,
$\theta$ is a phase, $\textbf{k}$ is the wave-number vector,
$\hat{a}^\dagger_{\textbf{k}\uparrow(\downarrow)}$ creates a fermion with
momentum $\textbf{k}$ and spin up (down), and $|0\rangle$ represents the
vacuum. However, this function is not an eigenstate of the particle
number operator. The BCS wave function projected to a fixed number $A$
of particles, half with spin up and the other half with spin down,
can be written as the antisymmetrized product \cite{bou88}
\begin{align}
\label{eq:BCS_anti}
\psi_{BCS}&(\textbf{R},S) = \mathcal{A}[ \phi(\textbf{r}_1,s_1, \textbf{r}_{2},s_2) \times \nonumber \\ 
& \times \phi(\textbf{r}_3,s_3,\textbf{r}_{4},s_4)\dots  \phi(\textbf{r}_{A-1},s_{A-1},\textbf{r}_{A},s_A)],
\end{align}
where $\textbf{R}$ is a vector containing the particle positions
$\textbf{r}_i$, $S$ stands for the spins $s_i$, and $\phi$ is the
pairing function. This wave function can be calculated efficiently as
a Slater determinant or a pfaffian~\cite{gan09}. 
The simulation is performed by using pairing orbitals constructed from the
functions of Eq.~(\ref{eq:sch_cyl}), instead of the plane waves typically employed
in a periodic box or in a harmonic trap~\cite{Carlson:2014}. 
The pairing orbitals are given by
\begin{eqnarray}
& & \phi(\textbf{r}_1,s_1,\textbf{r}_2,s_2) = \sum_{\textbf{k}} \frac{v_{\textbf{k}}}{u_{\textbf{k}}} \mathcal{N}_{mp}^2 J_m\left(\frac{j_{mp}\rho_1}{{\cal R}}\right)  \times \nonumber \\
& & J_{m}\left(\frac{j_{mp}\rho_2}{{\cal R}}\right)  \exp{\lbrace i \left[ k_z(z_1-z_2)+m (\varphi_1-\varphi_2) \right] \rbrace } \times \nonumber \\ 
& & \left[ \product{s_1 s_2}{\uparrow \downarrow} - \product{s_1 s_2}{ \downarrow \uparrow} \right],
\end{eqnarray}
where we have explicitly included the spin part to impose singlet pairing.

We also want to simulate systems that are not fully paired, by including
unpaired states using a superposition of free-particle solutions,
\begin{equation}
\label{eq:unpaired}
\Phi(\rho,\varphi,z) = \sum_{n,m,p} \nu_{nmp} \Psi_{nmp}(\rho,\varphi,z), 
\end{equation}
where the $\nu_{nmp}$ are variational parameters. For $q$ pairs and $o$ unpaired states, $A=2q+o$, we have
\begin{eqnarray}
\label{eq:BCS_unpaired}
& & \psi_{BCS}(\textbf{R},S) = \nonumber \\ 
& & \mathcal{A}[\phi(\textbf{r}_1,s_1, \textbf{r}_{2},s_2) \dots \phi(\textbf{r}_{2q-1},s_{2q-1},\textbf{r}_{2q},s_{2q}) \times \nonumber \\
& &  \Phi_{2q+1}(\textbf{r}_{2q+1}) \dots \Phi_{2q+o}(\textbf{r}_{2q+o})].
\end{eqnarray} 
as described in Ref.~\cite{car03}.

\subsection{Ground state}
\label{sect:ground}

For the ground state of fermions in a cylindrical container, we
use \eq{eq:BCS_anti}, or \eq{eq:BCS_unpaired} if we have unpaired
particles. The momentum vectors in the cylinder are quantized and the
system has a shell structure that depends on ${\cal R}$
and $\mathcal{L}$, see Eq.~(\ref{eq:eigenenergies}). We consider
$\alpha_{\textbf{k}} \equiv v_{\textbf{k}}/u_{\textbf{k}}$ variational parameters \cite{[{See Supplemental Material at }][{for the
variational parameters of Equations \ref{eq:unpaired}, \ref{eq:pair_gs},
\ref{eq:jastrow_one} and \ref{eq:pair_vortex} for the different system
sizes.}]table} and we assume the pairing wave functions to be
\begin{eqnarray}
\label{eq:pair_gs}
& & \phi(\textbf{r},\textbf{r}') = \tilde{\beta}(\textbf{r},\textbf{r}') + \sum_{I\leqslant I_C} \alpha_I \mathcal{N}_{m p}^2 J_m\left(\frac{j_{mp}\rho}{{\cal R}}\right)  
  \times   \nonumber \\
& & J_{m}\left(\frac{j_{mp}\rho'}{{\cal R}}\right) \exp{\lbrace i \left[ k_z(z-z')+m (\varphi-\varphi') \right] \rbrace },
 \end{eqnarray}
where we hereafter adopt primed indexes to denote spin-down particles, and unprimed ones to refer to spin-up particles, and we omit the spin
part. In \eq{eq:pair_gs}, $I_C$ is a cutoff shell number and we assume
that contributions of shells with $I>I_C$ are included through
the $\tilde{\beta}(\textbf{r},\textbf{r}')$ function, given by
\begin{equation}
\tilde{\beta}(\textbf{r},\textbf{r}')=
\begin{cases}
\mathcal{N}_{01}^2 J_0\left(\frac{j_{01}\rho}{{\cal R}}\right) J_0\left(\frac{j_{01}\rho'}{{\cal R}}\right) \times \nonumber \\ 
\times \beta(r)+\beta({\cal L}-r)-2\beta({\cal L}/2)
 &\text{for } r\leqslant {\cal L}/2\\
0 &\text{for } r > {\cal L}/2
\end{cases}
\end{equation}
and
\begin{eqnarray}
\beta(r) = [1+\gamma b r][1-e^{-cbr}]\frac{e^{-br}}{cbr},
\end{eqnarray}
where $r=|\textbf{r}-\textbf{r}'|$ and $b$, $c$, and $\gamma$ are
variational parameters.
This functional form of $\beta(r)$ describes the short-distance
correlation of particles with anti-parallel spins.
We consider $c=10$, $\gamma=5$, and $b$ = 0.5 $k_F$.

We also include a one-body Jastrow factor of the form
\begin{equation}
\label{eq:jastrow_one}
\chi(\rho_i) = \left( \frac{a}{\sqrt{2\pi\sigma^2}} \exp{ \left\{ \frac{(\rho_i -\bar{\rho})^2}{2\sigma^2} \right\} } + \nu \right)^\lambda ,
\end{equation}
where $a$, $\sigma$, $\bar{\rho}$, $\nu$, and $\lambda$ are variational
parameters. % and $\rho_i$ is the distance from the cylinder axis.
The
correlation between antiparallel spins is included in the two-body
Jastrow factor $f(r_{ij'})$ obtained from solutions of the two-body
Schroedinger equation
\begin{equation}
\left[ -\frac{1}{M} \nabla^2 + V(r) \right]f(r<d)=\lambda f(r<d),
\end{equation}
where $d$ is a variational parameter. The boundary conditions are
$f(r>d)=1$ and $f'(r=d)=0$ \cite{cow02}. The total wave function for
the ground state of fermions in a cylindrical container is then written as
\begin{equation}
\label{eq:wave_BCS}
\psi_0(\textbf{R})=\prod_k \chi(\rho_k) \prod_{i,j'} f(r_{ij'}) \psi_{BCS}(\textbf{R}).
\end{equation}

\subsection{Vortex line}
\label{sect:vortex}

The simulation of a vortex excitation is done by considering pairing
orbitals that are eigenstates of $L_z$ with eigenvalues $\pm \hbar$. This
is done by coupling single particle states of quantum numbers
$m$ and $(-m+1$). In this case we consider pairing orbitals of the form
\begin{eqnarray}
\label{eq:pair_vortex}
 & & \phi_V(\textbf{r},\textbf{r}') = \tilde{\beta}(\textbf{r},\textbf{r}') + \sum_{i=1}^{K} \tilde{\alpha}_i \mathcal{N}_{m;p} \mathcal{N}_{-m+1;p} \left\{  \right.  \nonumber \\
 & &  J_m\left(\frac{j_{mp}\rho}{{\cal R}}\right) J_{-m+1}\left(\frac{j_{-m+1;p}\rho'}{{\cal R}}\right) \times \nonumber \\
 & & \exp{\lbrace i \left[ k_z(z-z')+m\varphi+(-m+1)\varphi') \right] \rbrace } +  \nonumber \\
& & J_m\left(\frac{j_{mp}\rho'}{{\cal R}}\right)  J_{-m+1}\left(\frac{j_{-m+1;p}\rho}{{\cal R}}\right) \times   \nonumber \\
& & \left.  \exp{\lbrace i \left[ k_z(z'-z)+m\varphi'+(-m+1)\varphi) \right] \rbrace } \right\}, 
 \end{eqnarray}
where $K$ is the number of single-particle states with quantum numbers
$(n,m,p)$ being paired with $(-n,-m+1,p)$. Since we do not want to have
a winding number in the $z$ direction, we consider the quantum number
$n$ for a particle and the time-reversed $-n$ for the other. Also, the
largest contribution is assumed to be from states with the same quantum
number $p$ in the radial part.

\subsection{Diffusion Monte Carlo method}
\label{sect:qmc}

The Hamiltonian of our system is given by
\begin{equation}
H=-\frac{\hbar^2}{2M}\sum_i\nabla_i^2+\sum_{i<j} V(r_{ij}) \,.
\end{equation}
The diffusion Monte Carlo (DMC) method is used to extract the lowest energy state of $H$ from an
initial state $\psi_T$ obtained through variational
Monte Carlo (VMC) calculations. 
The propagation of the system in an imaginary time $\tau$ can formally be written as
\begin{equation}
\psi(\tau) = e^{-(H-E_T)\tau} \psi_T,
\end{equation}
where we introduce an energy offset $E_T$, used to control the
normalization of $\psi(\tau)$. 
 In the limit $\tau \to \infty$, only the lowest energy
component $\Phi_0$, not orthogonal to $\psi_T$, survives:
\begin{equation}
\lim_{\tau \to \infty} \psi(\tau) = \lim_{\tau \to \infty} e^{-(H-E_T)\tau} \psi_T = \Phi_0.
\end{equation}
The evolution in imaginary time is performed by solving the integral equation
\begin{equation}
\label{eq:integral}
\psi(\textbf{R},\tau) = \int d\textbf{R}' G(\textbf{R},\textbf{R}',\tau)\psi_T(\textbf{R}'),
\end{equation}
where $G(\textbf{R},\textbf{R}',\tau)$ is the Green's function of the
Hamiltonian, which contains a diffusion term, related to the kinetic
operator, and a branching term depending on the potential. The exact form
of $G(\textbf{R},\textbf{R}',\tau)$ is known only for very simple
cases, but it can be approximated in the limit of $\Delta \tau \rightarrow
0$. An importance sampled version of \eq{eq:integral} is then solved iteratively with a small time step,
for a large number of steps. The DMC method can only sample positive
distributions. A system formed by fermions has the so-called fermion
sign problem. We overcome this problem by using the usual fixed-node
approximation \cite{and75}. For a detailed description of the DMC
algorithm, the importance sample technique, and the fermion sign problem,
the reader is referred to the review in \olc{fou01} and references therein.

Note that the trial wave function $\Psi_T(\textbf{R})$ is used in two
different ways: as an approximation of the ground state in the VMC calculation and as an
importance function that also determines the nodal surface followed by the fixed-node
approximation.

The variational parameters for the pairing functions and two-body
Jastrow factor have been optimized using the stochastic reconfiguration method
\cite{cas04}. The parameters for the one-body term are chosen to maximize
the overlap of the density profile along the radial coordinate calculated 
using DMC and VMC.

\section{Results}
\label{sect:res}

In this section we present the results obtained with
the BCS wave function $\psi_0$ for fermions in a cylinder,
Eq.~(\ref{eq:wave_BCS}), and the results for the system with a vortex
line along the $z$ axis, $\psi_V$ using the pair orbitals of
Eq.~(\ref{eq:pair_vortex}). 
Expectation values of operators that do not commute
with the Hamiltonian, such as the density, can be calculated using a
combination of mixed and variational estimators,
\begin{eqnarray}
\label{eq:extrapolation1}
\bra{\Phi} \hat{S} \ket{\Phi} \approx 2 \bra{\Phi} \hat{S} \ket{\Psi_T} - \bra{\Psi_T} \hat{S} \ket{\Psi_T} \nonumber \\ 
 +\mathcal{O}\left[ (\Phi-\Psi_T)^2 \right].
\end{eqnarray}
Such combinations of VMC and DMC estimators are called extrapolated
estimators \cite{cep86}.

We have fixed the number density at $k_F^3/(3\pi^2)$, which is the density
of the free Fermi gas, and we have freedom to choose the radius ${\cal
R}$ and the height ${\cal L}$ of the cylinder. In most simulations we
set ${\cal L}=2{\cal R}$, so that the diameter is equal to the height
of the cylinder; we have verified that the latter choice does not
affect the results.

\subsection{Density profile}
\label{sec:density_profile}

The spatial distribution of the particles in the
cylinder was studied by calculating the density profile $\mathcal{D}(\rho)$ along the radial direction
$\rho$. The normalization is chosen so that
\begin{equation}
\int_V {\cal D}(\rho) dv =1,
\end{equation}
where the integral is over the volume $V=\pi{\cal R}^2{\cal L}$ of the cylinder.

\begin{figure}[!htb]
  \centering
  \includegraphics[angle=-90,width=\linewidth]{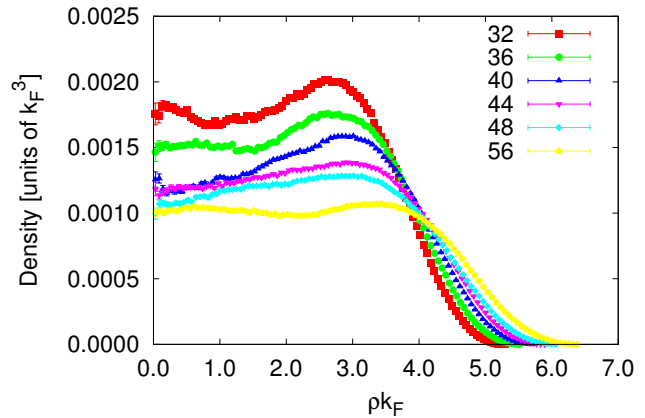}
  \caption{(Color online) Ground-state density profile for systems with closed shells,
  corresponding to different number of particles as indicated in the legend.}
  \label{fig:density_closed}
\end{figure}

The ground-state density profile for closed shells of the system is presented in Fig. \ref{fig:density_closed}. Boundary effects decrease as the number of particles considered is
increased. For the largest system the density has small fluctuations
near the center of the cylinder and it smoothly
decreases until it vanishes at the wall. The almost-constant density for small
$\rho$ is consistent with the ground state, since it corresponds to the
bulk of the system. For
the largest number of particles we have considered we assume size
effects to be negligible.

\begin{figure}[!htb]
  \centering
  \includegraphics[angle=-90,width=\linewidth]{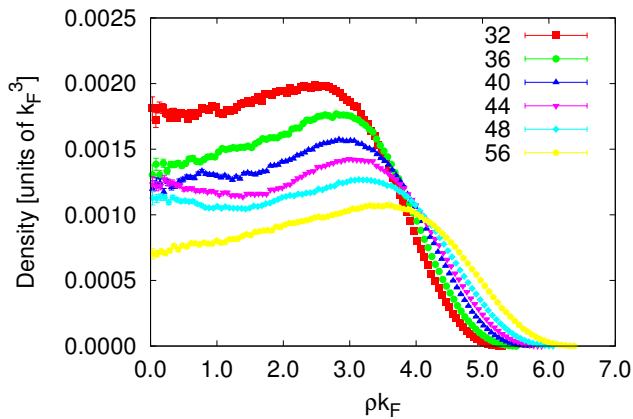}
  \caption{(Color online) Density profile for systems with a vortex line and closed shells for
  different numbers of particles.}
  \label{fig:density_vortex_closed}
\end{figure}

We present the density profile
for closed-shell systems with a vortex-line excitation in Fig. \ref{fig:density_vortex_closed}. The most
interesting feature of this quantity is the nonzero density at the core,
near $\rho=0$. Previous calculations using Bogoliubov-deGennes
theory \cite{bulgac2003,sensarma2006}, while showing a finite density at the
origin, give a much larger suppression of the density at the origin. Their
density at the origin at unitarity is about one-quarter of
the bulk density. We do not see such a large suppression. The reasons
for these differences could be due to our geometry, the fixed-node
approximation we use, or the approximations in the Bogoliubov-deGennes
theory. Future calculations using both methods in the same geometry
could help shed light on these differences.

We calculated the particle number a distance $R$
from the cylinder axis as
\begin{eqnarray}
\eta(R) =\int_0^L dz \int_0^{2\pi} d\varphi \int_0^R d\rho \ \rho \ \mathcal{D}(\rho).
\end{eqnarray}
We find that the difference in $\eta(R)$ for the ground state vs the vortex-line state is at most two particles. The optimization process is
computationally costly and it may be responsible for the difficulties
in resolving the densities of the two systems.

\subsection{Energy}
\label{sec:energy}

\subsubsection{Ground state}
\label{subsec:gs}

\begin{figure}[!htb]
  \centering
  \includegraphics[width=\linewidth]{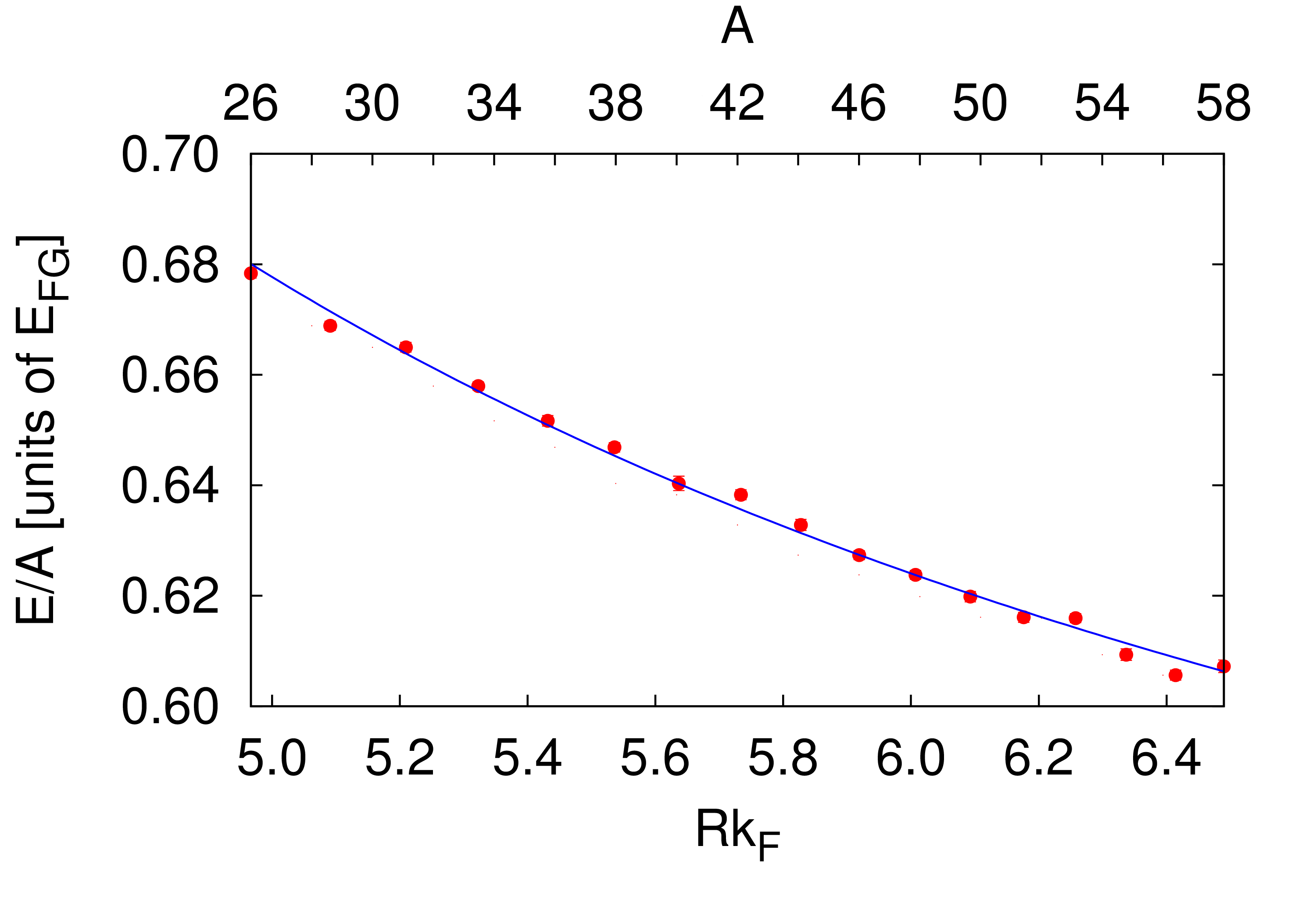}
  \caption{(Color online) Ground-state energy per particle for different system sizes. The solid line corresponds to the energy per particle as a function of $\mathcal{R}$, \eq{eq:fit}.}
  \label{fig:energy_gs}
\end{figure}

The energy per particle of the system in the cylindrical
geometry goes to the value of the bulk energy per particle in the limit of ${\cal R}, {\cal L} \to \infty$. Since the wave function vanishes at the cylinder walls of our
finite system, the energy has a dependence on the surface area of the
wall, $S=2\pi {\cal R L}=4 \pi {\cal R}^2$. However, we are still able
to estimate the bulk energy. We extrapolate the energy per particle
as a function of the radius using the functional form
\begin{equation}
\label{eq:fit}
\frac{E({\cal R})}{A} = E_0 + \frac{E_s}{4\pi{\cal R}^2},
\end{equation}
where $E_0$ and $E_s$ are constants, that represent
the bulk and surface energies. The resulting parameters are
$E_0=(0.50 \pm 0.01) E_{FG}$ and $E_s=(55.2
\pm 1.0) E_{FG} k_F^{-2}$, and $E({\cal R})$ is shown in
Fig. \ref{fig:energy_gs}. The $E_0$ parameter in this geometry
is analogous to the Bertsch parameter in a box with periodic
boundary conditions. The energy levels are much more degenerate
in the box compared to the cylinder. The translational invariance gives a good basis for plane waves, while Bessel functions are not as well defined for the radial direction,
which leads to a trial function with more 
parameters needed to simulate systems with the same
number of particles. For example, early QMC calculations in
the box 
\cite{car03} obtained an upper bound of the Bertsch
parameter $\xi=0.440(2)$ for $A=38$ using five parameters analogous to the 
$\alpha_I$ of \eq{eq:pair_gs}, so that the highest energy single particle state 
has $k_{\rm max}^2\approx 1.46 \ k_F^2$. If we consider the same number of 
particles and number density in the cylindrical geometry, we require 12 
$\alpha_I$'s to reach the same $k_{\rm max}^2$. This increased 
degeneracy may account for the higher values of
$E_0$ when compared to the upper bound of the Bertsch
parameter $\xi=0.383(1)$ \cite{gan11,Forbes:2011} and its exact
value \cite{car11}, $\xi=0.372(5)$.

These differences between the
periodic box simulations and the cylindrical simulations show that the
calculated
properties are significantly biased by the geometry. The clear
dependence on system size shown in Fig. \ref{fig:energy_gs} further
indicates that the main cause is due to nodal surface errors in our
fixed-node calculations.

We performed one simulation doubling the
height of the cylinder and the number of particles used in our
calculation with $A=26$. The energy per particle for this system is
$(0.683\pm0.001)E_{FG}$, which differs less than 1\% from the value found
for $A=26$, $(0.678\pm0.001)E_{FG}$, verifying that our results are independent of the condition
$\mathcal{L}=2\mathcal{R}$.

\subsubsection{Superfluid pairing gap}
\label{subsec:superfluid_pairing_gap}

\begin{figure}[!htb]
  \centering
  \includegraphics[angle=-90,width=\linewidth]{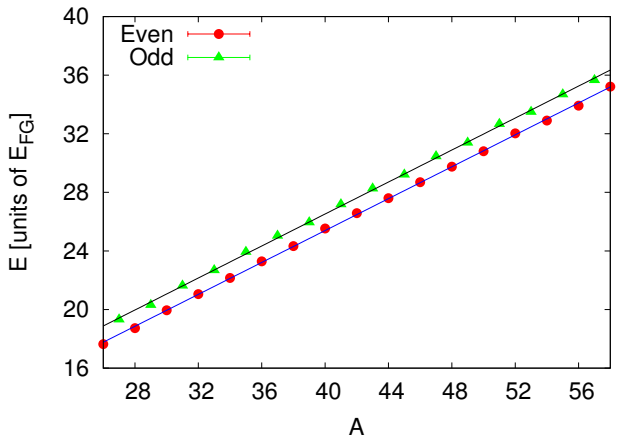}
  \caption{(Color online) Ground-state energy for even (circles) and odd (triangles) numbers of particles. Solid lines correspond to linear fits of the energy as a function of the number of particles for systems with even and odd numbers of particles.}
  \label{fig:gap}
\end{figure}

Experiments with cold atom gases determined the pairing gap to be
approximately half the Fermi energy \cite{shi07,car08}.  The pairing
gap at $T=0$ is calculated using the odd-even staggering formula
\cite{bro13},
\begin{equation}
\label{eq:staggering}
\Delta(A+1) = E(A+1)- \frac{1}{2}\left[ E(A)+E(A+2) \right].
\end{equation}
We consider that, for an even number of particles, all of them are
paired. For a system with an odd number of particles, the unpaired
particle is described by \eq{eq:unpaired}, and we take the
coefficients $\nu_{nmp}$ as variational parameters. We assume that $n$
and $m$ are good quantum numbers for the unpaired particle, because we
employ periodic boundary conditions in the $z$ direction and the wave
function must be an eigenstate of $L_z$. Thus, we chose the wave function
of the unpaired particle to be a linear combination of free-particle
states with the same $n$ and $m$, but different $p$, hence \eq{eq:unpaired}
reduces to a sum only over $p$. We perform independent simulations
for different values of $n$ and $m$ and we determine the $\{ \nu_{nmp}
\}$ which minimize the total energy of the system. In the calculation
of the gap we choose the unpaired orbital that gives the lowest energy. 
This is analogous to previous calculations in the bulk \cite{cha04,car05}.
In Fig. \ref{fig:gap} we show the total energy of the system for $26
\leqslant A \leqslant 58$. The pairing gap is estimated through
\eq{eq:staggering}, $\Delta=(1.12\pm 0.02)E_{FG}$. It is noteworthy
that the pairing gap is calculated using the difference in energy upper
bounds, thus the result is sensitive to the relative quality of the nodal
structure. It is likely that the optimization of the excited-state wave
function is less effective, which would overestimate the pairing gap.
We also note that finite-size effects might be canceled out in this calculation, which does not consider the vortex line.

\subsubsection{Excitation energy}

\begin{figure}[!htb]
  \centering
  \includegraphics[angle=-90,width=\linewidth]{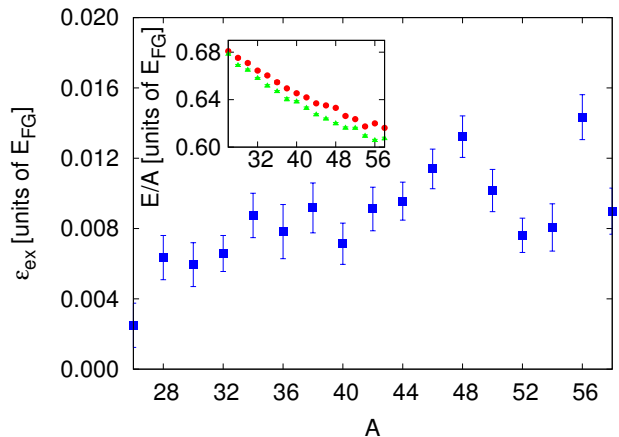}
  \caption{(Color online) Excitation energy per particle. Inset: Ground-state energies (triangles) and the energy of the system with a vortex line (circles).}
  \label{fig:exc}
\end{figure}

The excitation energy for a system with a vortex line in our
geometry is given by the
difference in energies between the excited state and the ground
state. In Fig. \ref{fig:exc} we present the excitation energy, as
well as the ground state and the system with a vortex-line energy, for
$26 \leqslant A \leqslant 58$. The average of the excitation energies per particle in
our geometry
for the larger systems ($42\leqslant A \leqslant 58$) is
$E_{ex}=(1.03 \pm 0.04)10^{-2}E_{FG}$.

\section{Discussion and conclusions}
\label{sect:con}

In this work, we have calculated the density profiles of the ground-state
and an excited state with a vortex line for a system of ultracold fermionic atoms 
at unitarity. For systems with $A \geqslant 36$ the ground state density profiles are flat near the center of
the cylinder and they smoothly decrease until the density vanishes at
the wall. The most interesting feature of the density profile of the
systems with a vortex line is the nonvanishing density at the core,
$\rho=0$. However, it is lower than the ground-state density by a small 
amount. Since the Cooper pairs have nonzero size, it is possible for
a pair to have nonzero angular momentum at the origin and still have
a nonzero density there.

For the cylindrical geometry, we calculated the energy of the ground
state for an even number of particles (all paired). Because the wave
function vanishes at the walls of the cylinder, we need very large values
of ${\cal R}$ and ${\cal L}$ to neglect the effects introduced by this
condition. We proposed a functional form for the energy per particle as
a function of the radius of the cylinder which takes into account the
energy term due to the walls.

The superfluid pairing gap of these ultracold atomic gases is of interest
because it is comparable to the Fermi energy of the system. The usual
odd-even staggering formula \cite{bro13} yields a gap of $\Delta=(1.12\pm
0.02)E_{FG}$. Previous quantum Monte Carlo simulations of fermions in
a box, using periodic boundary conditions, predicted $\Delta=(0.84 \pm
0.05)E_{FG}$ \cite{car05}, while an experiment at finite temperature
produced the value $\Delta=(0.45\pm0.05)E_{FG}$ \cite{car08}.

%The excitation energy is readily calculated from the energies of the
%ground state and the vortex line system. Associating a frequency $\omega$
%to this energy we can write
%\begin{equation}
%\label{eq:ex_omega}
%E_{ex} = \hbar \omega = (8.6\pm0.3)10^{-3}E_{FG}.
%\end{equation}
%We now compare our results to one of the milestones in the study
%of vortices in Fermi gases \cite{zwi05}, which used $^6$Li atoms to
%study the BCS-BEC crossover. The experiment focused on obtaining vortex
%lattices for different interaction strengths, ranging from the BCS
%to the BEC limits. In the experiment, it was found that the characteristic
%microscopic length was $1/k_F=0.3 \ \mu$m, which yields a Fermi energy
%$E_F=(\hbar^2 k_F^2)/(2m) \approx 3.30 \times 10^{-11}$ eV, where $m=1.17
%\times 10^{26}$ kg. If we use this Fermi energy in Eq. \ref{eq:ex_omega} we
%obtain $E_{ex}=\hbar \omega = (2.8\pm 0.1) \times 10^{-13}$
%eV. The rotational frequency of the lattice was found to be close
%to the stirring frequency, which is $f_{stirr} \simeq $ 45 Hz near
%unitarity. Hence, $\hbar \omega_{stirr} = \hbar\ 2\pi f_{stirr}\simeq$
%$1.9 \times 10^{-13}$ eV. Although we are comparing these two quantities,
%the systems are quite different: we studied a single vortex line, and
%the experiment has a vortex lattice. However, if we neglect the
%vortices' interactions in the experiment, the result is of the same
%order of magnitude.

Future calculations will include a more detailed study of the vortex
structure, the excitations of the fluid in the presence of a vortex,
and calculations of the reduced density matrices in order to better
understand the condensate in the presence of vortices.

We developed a wave function to study superfluidity and vortices in a
cylindrical geometry. This geometry enabled us to simulate a vortex line
in a superfluid Fermi gas using a bare Hamiltonian. These calculations
allowed theoretical predictions of the structure of vortices that can
be compared with experiments. Our results have implications both for
cold-atom research and for astrophysics, where the vortex structure in the
superfluid crust of neutron stars is not well understood.  This work can
be extended to study vortices in superfluid neutron matter by extending
the calculations in Refs. \cite{gan08} and \cite{gan09}.

\begin{acknowledgments}
This work was partially supported by Grants No. 10/10072-0, No.
12/24195-2, No. 13/19853-3, and No. 14/20864-2 from FAPESP and No. 087/2012 from
PVE/CAPES. The work of S.G. was supported by the U.S. Department of
Energy (DOE), Office of Nuclear Physics, under Contract No. DE-AC52-06NA25396,
and by the NUCLEI SciDAC program. 
K.S. was partially supported by National Science
Foundation Grant No. PHY-1404405. 
Computational resources were
provided by Los Alamos Open Supercomputing and
CENAPAD-SP at Unicamp.
We also used resources
provided by NERSC, which is supported by the U.S. DOE under Contract No.
DE-AC02-05CH11231. 

\end{acknowledgments}

\bibliography{article}% Produces the bibliography via BibTeX.

\end{document}